Fractional Method of Characteristics for Fractional Partial Differential Equations

Guo-cheng Wu*

Modern Textile Institute, Donghua University, 1882 Yan-an Xilu Road,

Shanghai 200051, PR China

Abstract

The method of characteristics has played a very important role in mathematical physics. Preciously, it was used to solve the initial value problem for partial differential equations of first order. In this paper, we propose a fractional method of characteristics and use it to solve some fractional partial differential equations.

Keywords: Modified Riemann-Liouville Derivative; Fractional Method of Characteristics; Fractional Partial Differential Equations

## 1 Introduction

In the past centuries, many methods of mathematical physics have been developed to solve the partial differential equations (PDEs) [1 - 2], among which the method of characteristics is an efficient technique for PDEs [3].

Fractional partial differential equations have attracted many researchers' interests. Then one of the questions may be naturally proposed: would it be possible to derive the exact solutions of partial differential equations (FPDEs) using a fractional method of characteristics? Recently, with the modified Riemann-Liouville derivative [4], G. Jumaire ever proposed a Lagrange characteristic method [5] which can solve the FPDEs

$$a(x,t)\frac{\partial^{\beta}u(x,t)}{\partial x^{\beta}}+b(x,t)\frac{\partial^{\alpha}u(x,t)}{\partial t^{\alpha}}=c(x,t),\ \ 0<\alpha,\beta\leq 1, \tag{1}$$

where $\alpha=\beta$. However, in most cases, the two fractional order parameters $\alpha$ and $\beta$ may not be equivalent.

In this letter, we present a more general fractional method of characteristics and consider the case $\alpha\neq\beta$.

---

* Corresponding author, E-mail address: wuguocheng2002@yahoo.com.cn (G.C. Wu).

## 2 Properties of Modified Riemann-Liouville derivative and Integral

Through this paper, we adopt the fractional derivative in modified Riemann-Liouville sense. The modified Riemann-Liouville derivative has been successfully applied in successfully applied in the probability calculus [6], fractional Laplace problems [7], fractional variational calculus [8] and fractional variational iteration method [9].

Firstly, we introduce the definitions and properties of the fractional calculation. Assume $f : R \to R : x \to f(x)$ denote a continuous (but not necessarily differentiable) function and let the partition $h > 0$ of an interval [0, 1]. Through the fractional Riemann Liouville integral

$$_0I_x^\alpha f(x) = \frac{1}{\Gamma(\alpha)}\int_0^x (x-\xi)^{\alpha-1} f(\xi)d\xi,\ \alpha > 0. \tag{2}$$

The modified Riemann-Liouville derivative is defined as

$$_0D_x^\alpha f(x) = \frac{1}{\Gamma(1-\alpha)}\frac{d}{dx}\int_0^x (x-\xi)^{-\alpha}(f(\xi)-f(0))d\xi, \tag{3}$$

where $x \in [0,1],\ 0 \le \alpha < 1.$ One also can obtained Eq. (3) through consequence of a more basic definition, a local one, in terms of a fractional finite difference [10]

$$\Delta^\alpha = (FW - 1)^\alpha f(x) = \sum_0^\infty (-1)^k \binom{\alpha}{k} f(x + (\alpha - k)h) \tag{4}$$

where $FWf(x) = f(x+h).$ Then the fractional derivative is defined as the following limit

$$f^{(\alpha)} = \lim_{h\to 0}\frac{\Delta^\alpha f(x)}{h^\alpha}. \tag{5}$$

Some properties for the proposed modified Riemann-Liouville derivative are listed

(a) Fractional Leibniz product law [10]

$$_0D_x^{(\alpha)}(uv) = u^{(\alpha)}v + uv^{(\alpha)}, \tag{6}$$

where $u^{(\alpha)} = {}_0D_x^{(\alpha)}(u)$. Much more generally, we can have

$$ {}_0D_x^{(n\alpha)}(uv)=\sum_{k=0}^{n}C_n^k u^{(k\alpha)}v^{((n-k)\alpha)}. \tag{7} $$

(b) Fractional Leibniz Formulation [10]

$$ {}_0I_x^{\alpha}D_x^{\alpha}f(x)=f(x)-f(0), 0<\alpha\le 1 \tag{8} $$

Note that the properties (a) and (b) lead to the integration by parts:

$$ {}_aI_b^{\alpha}u^{(\alpha)}v=(uv)/_a^b-{}_aI_b^{\alpha}uv^{(\alpha)}. \tag{9} $$

(c) Fractional Jumarie-Taylor series [10]

$$ f(x+h)=\sum_{k=0}^{\infty}\frac{h^{\alpha k}}{(\alpha k)!}f^{(\alpha k)}(x). \tag{10} $$

(d) Integration with respect to $(dx)^{\alpha}$

Assume $f(x)$ denote a continuous $R\to R$ function. We use the following equlitiy for the integral w.r.t $(dx)^{\alpha}$ [10]

$$ {}_0I_x^{\alpha}f(x)=\frac{1}{\Gamma(\alpha)}\int_0^x(x-\xi)^{\alpha-1}f(\xi)d\xi=\frac{1}{\Gamma(\alpha+1)}\int_0^x f(\xi)(d\xi)^{\alpha}, 0<\alpha\le 1. \tag{11} $$

(e) Some useful formulas

$$ \begin{aligned} &f([x(t)])^{(\alpha)}=\frac{df}{dx}x^{(\alpha)}(t);\\ &{}_0D_x^{(\alpha)}x^{\beta}=\frac{\Gamma(1+\beta)}{\Gamma(1+\beta-\alpha)}x^{\beta-\alpha};\\ &\int(dx)^{\beta}=x^{\beta};\\ &\Gamma(1+\alpha)dx=d^{\alpha}x. \end{aligned} \tag{12} $$

## 3 Fractional Method of Characteristics

It is well known that the method of characteristics has played a very important role in mathematical physics. Preciously, the method of characteristics is used to solve the initial value problem for general first order. Consider the following first order equation,

$$ a(x,t)\frac{\partial u(x,t)}{\partial x}+b(x,t)\frac{\partial u(x,t)}{\partial t}=c(x,t). \tag{13} $$

The goal of the method of characteristics is to change coordinates from $(x,t)$ to a new coordinate system $(x_0, s)$ in which the PDE becomes an ordinary differential equation (ODE) along certain curves in the $x-t$ plane. The cures are called the characteristic curves which read

$$\frac{du}{c(x,t)} = ds,$$
$$\frac{dx}{a(x,t)} = ds,$$
$$\frac{dt}{b(x,t)} = ds.$$

With the modified Riemann-Liouville derivative, G. Jumaire ever gave a Lagrange characteristic method [5] which can solve some classes of fractional partial differential equations. In this paper, we present a more generalized fractional method of characteristics and use it to solve linear fractional partial equations.

More generally, we extend this method to linear space-time fractional differential equations

$$a(x,t)\frac{\partial^{\beta} u(x,t)}{\partial x^{\beta}} + b(x,t)\frac{\partial^{\alpha} u(x,t)}{\partial t^{\alpha}} = c(x,t), 0<\alpha, \beta \le 1 \quad (14)$$

Expand $u$ as the fractional Jumarie-Taylor's series of multivariate functions [10],

$$du = \frac{\partial^{\beta} u(x,t)}{\Gamma(1+\beta)\partial x^{\beta}}(dx)^{\beta} + \frac{\partial^{\alpha} u(x,t)}{\Gamma(1+\alpha)\partial t^{\alpha}}(dt)^{\alpha},\ 0<\alpha,\beta\le 1. \quad (15)$$

The total derivative here is more generalized. The function $u$ here is $\beta^{th}$ order differentiable with respecter to $x$ and $\alpha^{th}$ order differentiable to $t$, respectively.

Similarly, note that the generalized characteristic curves can be presented by

$$\frac{du}{ds} = c(x,t),$$
$$\frac{(dx)^{\beta}}{\Gamma(1+\beta)ds} = a(x,t), \quad (16)$$
$$\frac{(dt)^{\alpha}}{\Gamma(1+\alpha)ds} = b(x,t).$$

Eq. (16) can be simplified as Jumaire's Lagrange method of characteristics (See. A. 12 in Ref. [5]) if we assume $\alpha=\beta$

$$\frac{(dx)^{\alpha}}{a(x,t)}=\frac{(dt)^{\alpha}}{b(x,t)}=\frac{\Gamma(1+\alpha)du}{c(x,t)}=\frac{d^{\alpha}u}{c(x,t)}. \tag{17}$$

Obviously, if $\alpha=1$ in Eq. (19), we can get the characteristic curve for Eq. (13)

$$a(x,t)\frac{\partial u(x,t)}{\partial x}+b(x,t)\frac{\partial u(x,t)}{\partial t}=c(x,t).$$

## 4 Application of Fractional Method of Characteristics

Example 1. As the first example, we consider space-time fractional equations for the transport equation in porous media,

$$\frac{\partial^{\alpha}u(x,t)}{\partial t^{\alpha}}+c\frac{\partial^{\beta}u(x,t)}{\partial x^{\beta}}=0,\ \ 0<\alpha,\beta\le 1. \tag{18}$$

Assume Eq. (20) subjects to the initial value $u(x,0)=\varphi(x)$.

The generalized characteristic curves satisfy

$$\begin{aligned}&\frac{du}{ds}=0,\\&(dx)^{\beta}=c\Gamma(1+\beta)ds,\\&(dt)^{\alpha}=\Gamma(1+\alpha)ds.\end{aligned} \tag{19}$$

Then we can obtain

$$\begin{aligned}&\frac{x^{\beta}}{\Gamma(1+\beta)}=cs+C_1,\\&\frac{t^{\alpha}}{\Gamma(1+\alpha)}=s+C_2,\\&u=C_3.\end{aligned} \tag{20}$$

where $C_1$, $C_2$ and $C_3$ are integral constants. Eliminating the parameter $s$, we find that the fractional curves $\frac{x^{\beta}}{\Gamma(1+\beta)}-\frac{ct^{\alpha}}{\Gamma(1+\alpha)}=x_0$ and $u$ is a constant along the fractional curves.

Then we can directly derive the exact solution of Eq. (18) has the following form

$$u = f\left(\frac{x^{\beta}}{\Gamma(1+\beta)} - \frac{ct^{\alpha}}{\Gamma(1+\alpha)}\right), \tag{21}$$

where $f\left(\frac{x^{\beta}}{\Gamma(1+\beta)}\right) = \varphi(x)$. By setting $u(x,t) = f\left(\frac{x^{\beta}}{\Gamma(1+\beta)} - \frac{ct^{\alpha}}{\Gamma(1+\alpha)}\right)$, we directly have an exact solution for the initial-value problem here.

Example. 2. As the second example, we investigate the more complicated equation,

$$\frac{t^{\alpha}}{\Gamma(1+\alpha)} \frac{\partial^{\alpha} u(x,t)}{\partial t^{\alpha}} + \frac{x^{\beta}}{\Gamma(1+\beta)} \frac{\partial^{\beta} u(x,t)}{\partial x^{\beta}} = 0, \ \ 0 < \alpha, \beta \le 1. \tag{22}$$

We can have the generalized cure equations

$$\begin{aligned} &\frac{du}{ds} = 0, \\ &\frac{(dx)^{\beta}}{\Gamma(1+\beta)ds} = \frac{x^{\beta}}{\Gamma(1+\beta)}, \\ &\frac{(dt)^{\alpha}}{\Gamma(1+\alpha)ds} = \frac{t^{\alpha}}{\Gamma(1+\alpha)}. \end{aligned} \tag{23}$$

With the properties (12), direct calculation leads to

$$\begin{aligned} &\frac{x^{\beta}}{\Gamma(1+\beta)} = c_1 e^{s}, \\ &\frac{t^{\alpha}}{\Gamma(1+\alpha)} = c_2 e^{s}, \\ &u = c_3. \end{aligned} \tag{24}$$

Then we can find Eq. (22) has the following solutions

$$u = f\left(\frac{x^{\beta}}{\Gamma(1+\beta)} / \frac{t^{\alpha}}{\Gamma(1+\alpha)}\right), \tag{25}$$

where the function $f$ is arbitrary .

Assume $X = \frac{x^{\beta}}{\Gamma(1+\beta)} / \frac{t^{\alpha}}{\Gamma(1+\alpha)}$. With the properties (12), we note that

$$\frac{\partial^{\alpha} u(x,t)}{\partial t^{\alpha}} = f_X X_t^{(\alpha)} = -f_X \frac{x^{\beta}}{\Gamma(1+\beta)} / \frac{t^{2\alpha}}{(\Gamma(1+\alpha))^2}, \tag{26}$$

and

$$\frac{\partial^{\beta} u(x,t)}{\partial x^{\beta}} = f_X X_x^{(\beta)} = f_X / \frac{t^{\alpha}}{\Gamma(1+\alpha)}. \tag{29}$$

As a result, we can proof

$$\begin{aligned}
&\frac{t^{\alpha}}{\Gamma(1+\alpha)} \frac{\partial^{\alpha} u(x,t)}{\partial t^{\alpha}} + \frac{x^{\beta}}{\Gamma(1+\beta)} \frac{\partial^{\beta} u(x,t)}{\partial x^{\beta}} \\
&= f_X \frac{x^{\beta}}{\Gamma(1+\beta)} / \frac{t^{\alpha}}{\Gamma(1+\alpha)} - f_X \frac{x^{\beta}}{\Gamma(1+\beta)} / \frac{t^{\alpha}}{\Gamma(1+\alpha)} \\
&= 0.
\end{aligned}$$

## 5. Conclusion

The classical method of characteristics is an efficient technique for solving partial differential equations. In this paper, fractional method of characteristics is under consideration for some classes of fractional partial differential equations and two examples are illustrated to show its efficiency. Besides, the presented method provides a potential tool to solve fractional symmetry equations in Lie group method. We will discuss such work in future.

## 6. Acknowledgement

The author feels grateful to Prof. G. Jumarie (Department of Mathematics, University of Quebec at Montreal, Canada) for his sincere help in preparing this paper.